\documentclass[prl,twocolumn,showpacs,floatfix]{revtex4}

\usepackage{amssymb}
\usepackage{graphicx}
\usepackage{amsmath}
\usepackage{bm}

\def\CC{{\rm\kern.24em \vrule width.04em height1.46ex depth-.07ex \kern-.30em C}}
\def\RR{{\rm\kern.24em \vrule width.04em height1.46ex depth-.07ex
\kern-.30em R}}
\def\P{{\rm I\kern-.25em P}}
\newcommand{\ket}[1]{\left \vert #1 \right \rangle}

\begin{document}

\title{Quantum tensor product structures are observable-induced}
\author{Paolo Zanardi$^{1,2,3}$, Daniel A. Lidar$^4$ and Seth Lloyd$^1$}
\affiliation{$^1$Department of Mechanical Engineering, Massachusetts Institute of
Technology, Cambridge, MA 02139}
\affiliation{$^2$Institute for Scientific Interchange (ISI) Foundation, Viale Settimio
Severo 65, I-10133 Torino, Italy}
\affiliation{$^3$Instituto Nazionale per la Fisica della Materia (INFM)}
\affiliation{$^4$Chemical Physics Theory Group, Chemistry Department, University of Toronto, 80 St. George St., Toronto,
ON M5S 3H6, Canada}

\begin{abstract}
It is argued that the partition of a quantum system into subsystems is
dictated by the set of operationally accessible interactions and measurements. The
emergence of a multi-partite tensor product structure of the
state-space and the associated notion of quantum entanglement are then
relative and observable-induced. We develop a general algebraic
framework aimed to formalize this concept. We discuss several cases
relevant to quantum information processing and decoherence control.
\end{abstract}

\pacs{PACS numbers: 03.67.Lx, 03.65.Fd, 03.65.Ud}
\maketitle

Suppose one is given a four-state quantum system. How does one decide
whether such a system supports entanglement or not? In other words,
should the given Hilbert space ($\CC^4$) be viewed as bi-partite
($\cong \CC^2 \otimes \CC^2$), or irreducible? In the former case
there exists a \emph{tensor product structure} (TPS) that supports
two entanglable qubits. In this case one finds a sharp dichotomy
between the quantum and classical realms, as perhaps most dramatically
exemplified in quantum information processing \cite{BD}. In the
irreducible case there is no entanglement and hence none of the
advantages associated with \emph{efficient} quantum information
processing \cite{Ekert,Meyer}.

Here we propose that \emph{a partitioning of a given Hilbert space is}
induced \emph{by the experimentally accessible observables
(interactions and measurements)} (see also
Refs.~\cite{virt,Barnum,VKL-WL}). Thus it is meaningless to refer to a
state such as the Bell state $\ket{\Phi ^{+}} =(\ket{0} \otimes
\ket{0} + \ket{1} \otimes \ket{1})/\sqrt{2}$ as entangled
\cite{entang}, without specifying the manner in which one can
manipulate and probe its constituent physical degrees of freedom. In
this sense \emph{entanglement is always relative to a particular set
of experimental capabilities}. Before introducing a formalization, let
us illustrate these ideas by means of a simple example.

\textit{Example 0: Bell basis}.--- Let $\ket{x} \otimes \ket{y} \equiv
\ket{x,y}$, $
(x,y\in \{0,1\})$ be the standard product basis for a two-qubit
system. Each qubit forms a subsystem. With respect to (wrt) this
bi-partition the Bell-basis states $\ket{\Phi ^{\pm }} =(\ket{00} \pm
\ket{11})/\sqrt{2}$, $\ket{\Psi^{\pm }} =(\ket{01} \pm
\ket{10})/\sqrt{2}$, are maximally-entangled. Now note that these can
be rewritten as $\ket{\chi^{\lambda }} :=\ket{\chi} \otimes
\ket{\lambda} $, where $\chi =\Phi ,\Psi $ and $\lambda =+,-$. With
respect to this new bi-partition the Bell states are by definition
\emph{product} states, and the subsystems are the $\chi $ and $\lambda
$ degrees of freedom. On the other hand the states $ \ket{x,y} $ are
now entangled and can be used for entanglement-based quantum
information protocols such as teleportation \cite{BD}. This striking
difference can be highlighted by considering the \textsc{swap}
operator $S$, which is non-entangling in the usual $x,y$-bipartition,
but, in the $\chi,\lambda $-bipartition one has
$S\ket{\chi,\lambda}=(-1)^{\chi \lambda}\ket{\chi,\lambda}$. Thus $S$
realizes a controlled phase-shift over $\ket{11} := \ket{\Psi ^{-}}$,
and in the new decomposition \textsc{swap} is a maximally-entangling
operator. Which then is the correct characterization of the TPS and
the associated entanglement? The answer depends on the set of
accessible interactions and measurements. In stating that the Bell states are entangled
one is implicitly assuming that there is experimental access to
(local) observables of the form $\{\sigma ^{\alpha }\otimes
\openone\},\{\openone\otimes \sigma ^{\beta }\}$ (where $ \alpha
,\beta \in \{x,y,z\}$ and $\sigma $ are the Pauli matrices). But this
assumption may not always be justified. For example, in quantum dot
quantum computing proposals utilizing electron spins \cite{LD}, it is
more convenient to manipulate exchange interactions than to control
single spins \cite{BKLW-Div-ZLc,LW}. In such cases the
accessible interactions may be non-local, and this is precisely the
situation that favors the $\chi ,\lambda $-bipartition, that then
acquires the same operational status as the standard $x,y$ one.
%We shall return to this example below, and exhibit the relevant interactions.

\textit{General framework}.--- We now lay down a conceptual framework
aimed to capture in its generality and relativity the notion of
``induced tensoriality'' of subsystems. Our definitions will be
observable-based and will mostly involve algebraic objects \cite{not}.
Let us consider a quantum system with finite-dimensional state-space $
\mathcal{H }$, a subspace $\mathcal{C}\subseteq \mathcal{H}$, and a
collection $\{\mathcal{A} _{i}\}_{i=1}^{n}$ of subalgebras of
End$(\mathcal{C })$ satisfying the following three axioms:

i) \emph{Local accessibility}: Each $\mathcal{A}_{i}$ corresponds to a
set of controllable observables.

ii) \emph{Subsystem independence}:
$[\mathcal{A}_{i},\,\mathcal{A}_{j}]=0$ $(\forall i\neq j)$.

iii) \emph{Completeness}: $\vee _{i=1}^{n}\mathcal{A}_{i}\cong \otimes
_{i=1}^{n} \mathcal{A}_{i}\cong \mathrm{End}(\mathcal{C})$.

Notice that the standard case of $N$ qudits ($d$-level systems)
$\mathcal{C} = \mathcal{H}=(\CC^{d})^{\otimes \,N}$ is the case
$\mathcal{A}_{i}\cong M_{d}$ $\forall i$ acting as the identity over
all factors (subsystems) but the $i$th one. Now we discuss the
physical meaning of the axioms i)--iii).

Axiom i) simply defines the basic algebraic objects at our
disposal. These objects are controllable observables (Hamiltonians
with tunable parameters, measurements).

Axiom ii) addresses separability. In order to claim that a system is
composite it must be possible to perform operations manipulating a
well-defined set of degrees of freedom \emph{while leaving all the
others unaffected}. Typically this is achieved by having individually
addressable, \emph{spatially} separated subsystems $i$ (e.g., a single
excess electron per quantum dots \cite{LD}), but as we shall see this
is certainly not the only possibility.
%% One can also view Axiom ii) as the well-known mathematical way
%% to encode in the quantum kinematics the dynamical independence of
%% (equal-time) operators associated to different physical degrees of freedom.
%% In field theory causal separation of local field supports is a sufficient
%% way to achieve such a condition \cite{haag}.

Axiom iii) is the crucial one in order to ensure that our
observable-based definition of multi-partiteness induces a
corresponding one at the state-space level. Its meaning will follow from Proposition 1 below: all the
operations not affecting the state of a subsystem (its symmetries),
are realized by operators corresponding to non-trivial operations only
over the degrees of freedom of the \emph{other} subsystems. All
symmetries are then \emph{physical operations} and no superselection
rules \cite{giulini,ss-qip} are present when a suitable state space $\mathcal{C}$ is chosen. When $\mathcal{C}$ is a proper subspace of $
\mathcal{H}$, we are dealing with an ``encoding'', a notion that has
proved useful, e.g., in quantum error correction and avoidance
\cite{ERR,EAC,KLV} and encoded universality \cite{BKLW-Div-ZLc,LW}.
%% Note that a generic operator over $\mathcal{H}$ which does
%% \emph{not} belong to End$(\mathcal{C})$, will in general induce
%% \emph{leakage} out of $\mathcal{C}$.
Generalizing Ref.~\cite{virt} we have the following central result:
%% , which shows the relation of the algebraic axioms i)--iii) to a
%% state-space picture:

\emph{Proposition 1}. A set of subalgebras $\mathcal{A}_{i}$
satisfying Axioms i)--iii) induces a TPS $\mathcal{C}=\otimes
_{i=1}^{n}\mathcal{H} _{i}$. We call such a multi-partition an
\emph{induced TPS}.

The proof is given in \cite{proof}.

We proceed to explore some consequences of the notion of observable-induced TPSs.  We first
briefly return to the example of Bell-states discussed above, then
continue the discussion at a more general level, and illustrate with
examples of unusual and dynamic TPSs.

\textit{Example 1}.--- Assume that one is given the following set of
independently controllable two-body interactions $\{\sigma ^{y}\otimes
\sigma ^{z},\sigma ^{z}\otimes \sigma ^{z},\sigma ^{x}\otimes \sigma
^{y},\sigma ^{x}\otimes \sigma ^{x}\}$. These interactions generate
the following subalgebras: $\mathcal{A}_{\chi }:=\{\openone,\sigma
^{x}\otimes \openone,\sigma ^{y}\otimes \sigma ^{z},\sigma ^{z}\otimes
\sigma ^{z}\}$, $ \mathcal{A}_{\lambda }:=\{\openone,\openone\otimes
\sigma ^{z},\sigma ^{x}\otimes \sigma ^{y},\sigma ^{x}\otimes \sigma
^{x}\}$. These satisfy Axioms i)-iii) (with $\mathcal{C}=\CC^{4}$) and
act, respectively, as local identity and Pauli $x,y,z$ matrices on the
$ \chi $ and $\lambda $ degrees of freedom considered above. Thus by
Prop.~1 $ \mathcal{A}_{\chi }$ and $\mathcal{A}_{\lambda }$ induce a
TPS $\CC^4 \cong \CC^{2}\otimes \CC^{2}$, namely, the $\chi ,\lambda $
bi-partition.

\textit{Superselection}.--- An important example for which one is led
to consider non-standard TPSs is systems exhibiting
\emph{superselection} rules \cite{giulini}. There the only allowed
physical operations correspond to operators commuting with a set of
superselection charges $\{\mathcal{Q}_{l}\}_{l=1}^{M}$, e.g.,
particle-numbers, which generate an abelian algebra
$\mathcal{Q}$. Denoting by $\Pi _{\mathcal{Q}}$ the projector over the
commutant of $\mathcal{Q}$, the physically realizable subsystem
operations are $\Pi _{\mathcal{Q}}(\mathcal{A}_{i})$
($i=1,\ldots,n$). These projected algebras typically either a) define
a new invariant subspace $\mathcal{C}^{\prime }$ with a new induced
TPS, or b) do not satisfy axioms ii),iii) anymore and therefore fail
to induce a proper TPS. The associated notion of entanglement and
entanglement-based protocols then must be reconsidered \cite{ss-qip}.

\textit{Irreducible representations}.--- A prototypical way for
obtaining an encoded bi-partite TPS is to consider the decomposition
of $\mathcal{H}$ into irreducible representations (irreps) of a
*-subalgebra $\mathcal{A}$ \cite{virt}. In that case
\begin{equation}
  \mathcal{H}\cong \oplus _{J}\CC
  ^{n_{J}}\otimes \mathcal{H}_{J},
  \label{eq:decomp}
\end{equation}
where the $\mathcal{H}_{J}$ are the $d_{J}$-dimensional irreps of
$\mathcal{A}$ and $n_{J}$ their multiplicities. The algebra
(commutant) can then be written as $\mathcal{A}\cong \oplus _{J}
\openone _{n_{J}}\otimes M_{d_{J}}$ ($\mathcal{A}^{\prime }\cong
\oplus_{J}M_{n_{J}}\otimes \openone_{d_{J}}$) \cite{KLV}.
%%% \cite{ALG}.
Upon restriction to a particular $J$-sector one has $\mathcal{A}\vee
\mathcal{A}^{\prime }\cong M_{n_{J}}\otimes M_{d_{J}}\cong
\mathcal{A}\otimes \mathcal{A}^{\prime }$. Then, according to Prop. 1,
$\mathcal{A}$ and $\mathcal{A}^{\prime }$ induce an (encoded)
bi-partite TPS in each irreducible block.

\textit{Example 2: Encoded tensoriality}.--- As an example of the
above construction, let $\mathcal{H}_{N}:=(\CC ^{2})^{\otimes \,N}$
denote an $N$-qubit space, $\mathcal{A}_{1}$ the algebra of totally
symmetric operators in End$(\mathcal{H}_{N})$, and $ \mathcal{A}_{2}$
the algebra of permutations exchanging the qubits. $\mathcal{A}_{1}$ is generated by the \emph{collective} spin
operators, i.e., $\mathcal{A}_{1}=\CC\{S^{\alpha
}:=\sum_{i=1}^{N}\sigma _{i}^{\alpha }\,|\,\alpha =x,y,z\}$, and
$\mathcal{A}_{2}=\mathcal{A} _{1}^{\prime }$ is generated by
Heisenberg exchange interactions: $\mathcal{A }_{2}=\CC\{\bm{\sigma
}_{i}\cdot \bm{\sigma }_{j}\}$ [$\bm{\sigma }=(\sigma ^{x},\sigma
^{y},\sigma ^{z})$]. In the context of decoherence-free subspaces and
subsystems \cite{EAC,KLV} $\mathcal{A}_{1}$ is the algebra of error
operators (system-bath interactions) and $\mathcal{A}_{2}$ is the
algebra of allowed quantum computational operations. Here our
perspective is quite different: we view both as algebras of
accessible interactions that induce a TPS. This is in fact an
\emph{encoded} TPSs, since one has (for even $N$) the
Hilbert space decomposition (\ref{eq:decomp}) with $J=0,...,N/2$,
${\cal H}_J=\CC^{d_J}$, $d_J=2J+1$, and
$n_{J}(N)=(2J+1)N!/[(N/2+J+1)!(N/2-J)!]$. Each summand in
Eq.~(\ref{eq:decomp}) is a code subspace with a bi-partite TPS.  We
stress the unusual feature of this example: the two ``qudits'' (i.e.,
subsystems) comprising the TPS need not have the same dimension
(though they do for $J=N/2-1$), and are manipulable by interactions of
a \emph{physically distinct} nature. The left (right) qudit is
manipulated by tuning only Heisenberg exchange couplings (global
magnetic fields). This example therefore has implications for
spin-based quantum computation \cite{LD}, where single-spin addressing
is technically very demanding.
%% In order to enact full control over the ``magnetic'' (right) qudit
%% one should take the $J=1/2$ irrep, since then $2J+1=2$ and global
%% magnetic fields can generate su$(2)$. Finally, note that this
%% situation is quite different from the paradigm of Heisenberg-only
%% quantum computation \cite{BKLW-Div-ZLc,LW}, where one in effect
%% discards the ``magnetic'' qudit.

\textit{Nested subalgebra chains}.--- The commutant construction
illustrated above provides a general way to realized an encoded
\emph{bi}-partite TPS.  In order to obtain encoded TPSs with more than
two subsystems we consider a \emph{nested chain of subalgebras}:
\begin{equation}
  \mathcal{B}_{0}\supset \mathcal{B}_{1}\supset \cdots \supset \mathcal{B}_{n}.
\label{chain}
\end{equation}
We assume that $\mathcal{B}_{0}$ acts irreducibly over
$\mathcal{H}$. Then $ \mathcal{H}$ typically will be reducible wrt
$\mathcal{B}_{i\geq 1}$. In particular, wrt $ \mathcal{B}_{2}$:
$\CC^{d_{J_{1}}}\cong \oplus _{J_{2}}\CC ^{n_{J_{2}}}\otimes
\CC^{d_{J_{2}}}$ and $\mathcal{B}_{2}\cong \oplus
_{J_{1},J_{2}}\openone_{n_{J_{1}}}\otimes \openone_{n_{J_{2}}}\otimes
M_{d_{J_{2}}}$. By iterating over the subalgebra chain one obtains:
\begin{equation}
  \mathcal{H}\cong \oplus _{J_{1},\ldots ,J_{n}}\otimes
  _{k=1}^{n}\CC^{n_{J_{k}}}\otimes \CC^{d_{J_{n}}}.
\label{chain-split}
\end{equation}
\emph{This is a sum over code subspaces
}$H(J_{1},\ldots ,J_{n}):=\otimes _{k=1}^{n}\CC^{n_{J_{k}}}\otimes
\CC^{d_{J_{n}}}$ \emph{with a multi-partite TPS}.
%\cite{ind}.
The nontrivial ones are those for which at least one
$n_{J_{k}}>1$. Note that while $\mathcal{B}_{2}$ has non-trivial
action only on $\CC^{d_{J_{2}}}$, $ \mathcal{B}_{1}$ has non-trivial
action on $\CC^{d_{J_{1}}}\supset \CC ^{d_{J_{2}}}$. So how does one
operate on a particular subsystem (qudit), say $\CC^{n_{J_{k}}}$?
% I.e., what are the corresponding commuting subalgebras that allow
% one to address each qudit?
We come to our second main result:

\emph{Proposition 2}. Given a nested subalgebra chain as in
Eqs.~(\ref{chain}),(\ref{chain-split}), the subsystems algebras are
given by
\begin{equation}
 \mathcal{A}_{i}=\mathcal{B}_{i}^{\prime }\cap \mathcal{B}_{i-1},\qquad
 (i=1,\ldots ,n).
\end{equation}
Conversely, when a set of subsystem algebras
$\{\mathcal{A}_{i}\}_{i=1}^{n}$ is given, the nested chain
$\mathcal{B}_{i}:=\vee _{k=i+1}^{n}\mathcal{A} _{k} $, $(i=1,\ldots
,n)$ results.

The proof is given in \cite{proof2}. We now illustrate the notion of
a nested subalgebra chain induced-TPS.

\textit{Example 3: Standard TPS}.--- The standard qubit-TPS over
$\mathcal{H} _{N}$ corresponds to the chain
$\mathcal{B}_{i}=\openone^{2^{i}}\otimes M_{2^{n-i}}$ $\,(i=1,\ldots
,n)$. In this case all the subalgebras are \emph{factors}, whence one
has a single $\mathcal{H}(J_{1},\ldots ,J_{n})$ term in
Eq. (\ref{chain-split}), with multiplicities $n_{J_{i}}=2$ and
dimensions $d_{J_{i}}=2^{n-i}$.
% (The quantum numbers $J_{i}$ are now simply indices and their
% numerical values are arbitrary).

\textit{Example 4: Stabilizer codes}.--- Consider $N$ qubits and the
following chain of nested algebras: $\mathcal{B}_{0}$ acts irreducibly
on $( \CC^{2})^{\otimes \,N}$; $\mathcal{B}_{1}$ acts trivially on the
first qubit but irreducibly on the rest, etc. To realize such a chain
let $ \{X_{1},\ldots ,X_{k}\}$ be a set of $N$-qubit, mutually
commuting operators, and let $\mathcal{B}_{i}=[\CC\{X_{1},\ldots
\,X_{i}\}]^{\prime }$, $(i=1,\ldots ,k)$. Further assume that the
$X_{i}$ are unitary, traceless, and square to the identity. Then the
corresponding Hilbert space decomposition is $\mathcal{H}\cong
(\CC^{2})^{\otimes \,i}\otimes (\CC ^{2})^{\otimes \,(n-i)}$, where
the first $i$ $\CC^{2}$ factors correspond to the $2^{i}$ possible
eigenvalues of $X_{1},\ldots ,X_{i}\in \mathcal{B} _{i}^{\prime
}$. When the $X_{i}$'s are generators of an abelian subgroup of the
Pauli group one recovers the stabilizer codes of quantum error
correction \cite{ERR}.

\textit{Example 5: Multi-partite encoded TPS}.--- Let us revisit Ex.~2
and show how a {\em multi}-partite encoded TPS is induced. Consider
$N=n\,2^{K}$ qubits, and the chain $\mathcal{B}_{0}:=
\CC\mathcal{S}_{N}$,
$\mathcal{B}_{i}:=\CC(\mathcal{S}_{N/2^{i}})^{\times 2^{i}}$,
$i=1,...,K$, where $\mathcal{S}$ denotes the symmetric group.
Conceptually, we have $2^{K}$ blocks of $n$ qubits each, and the
subalgebra chain corresponds to operating on these blocks with
increasing levels of resolution. By Prop. 2 we should find a
$K+1$-partite encoded TPS. To see this, recall that the state-space
$\mathcal{H}_{N}\cong $ $( \CC^{2})^{\otimes N}$ of $N$ qubits splits
wrt $\mathcal{S}_{N}$ exactly as in the su$(2)$ case (Ex.~2) except
that by the duality between $\mathcal{S}_{N}$ and su$(2)$, the role of
$n_{J}$ and $d_{J}$ is interchanged, while $J$ remains an su$(2)$
irrep label. E.g., for $N=6$ ($K=1$ and $n=3$) we have
$\mathcal{H}_{6}\cong \oplus _{J=0}^{3} \CC^{\tilde{n}_{J}}\otimes
\CC^{\tilde{d}_{J}}\cong H_{0}\otimes \CC ^{5}\oplus H_{1}\otimes
\CC^{9}\oplus H_{2}\otimes \CC^{5}\oplus H_{3}\otimes \CC$, where now
$\tilde{n}_{J}=2J+1$, $\tilde{d}_{J}=n_{J}(6)$, and
$H_{J}:=\CC^{2J+1}$, $J=0,1,2,3$. The chain then consists of
$\mathcal{B} _{0}=\CC\mathcal{S}_{6}$ and
$\mathcal{B}_{1}:=\CC(\mathcal{S}_{3}\times \mathcal{S}_{3})$, i.e.,
exchanges between the first three $\times $ second three qubits. From
Prop.~2 this algebra chain defines the encoded TPSs with algebra
subsystems given by $\mathcal{A}_{0}:= \mathcal{B}_{0}^{\prime }=$
totally symmetric operators (recall Ex.~2) and
$\mathcal{A}_{1}=\mathcal{B}_{1}^{\prime }\cap \mathcal{B}_{0}$, where
$ \mathcal{B}_{1}^{\prime }$ are block-symmetric operators, so that
$\mathcal{A }_{1}=$linear combination of permutations, symmetrized wrt
$\mathcal{S} _{3}\times \mathcal{S}_{3}$. Decomposing the
$\CC^{\tilde{d}_{J}}$ factors wrt $\mathcal{S}_{3}\times
\mathcal{S}_{3}$ we find, e.g., for the $H_1 \otimes \CC^9$ term that
it describes a qubit times a qutrit \cite{calc}. The operations over
the qutrit are provided by the algebra of totally symmetric six-qubits
operators. Those over the qubit are realized by operators in
$\CC\mathcal{S}_{6}$ having the form $X\otimes Y$ where $X$ ($Y$) is a
totally symmetric operators over the first (second) three qubits. For
example, elements of the form $\bm{\sigma }_{1+3i}\cdot \bm{\sigma
}_{2+3i}+ \bm{\sigma }_{2+3i}\cdot \bm{\sigma }_{3+3i}+\bm{\sigma
}_{3+3i}\cdot \bm{\sigma }_{1+3i},\,(i=0,1)$ have trivial action over
the qutrit (being a combination of $\mathcal{S}_{6}$ permutations) and
a non-trivial one over the qubit [being $su_{1-3}(2)\times
su_{4-6}(2)$ elements].

Returning to the case of $K$ blocks, one can see how an encoded
multi-partite TPS will emerge. For example, with $n=3$ and $K=2$ we
have the chain $\mathcal{B}_{0}=\CC\mathcal{S}_{12}\supset
\mathcal{B}_{1}:=\CC( \mathcal{S}_{6}\times \mathcal{S}_{6})\supset
\mathcal{B}_{2}:=\CC(\mathcal{S }_{3}\times \mathcal{S}_{3}\times
\mathcal{S}_{3}\times \mathcal{S}_{3})$.  By comparing, as in
\cite{calc}, the decompositions of $\mathcal{H}_{12}$ wrt $
\mathcal{B}_{1}$ and $\mathcal{B}_{2}$ one can identify the
tri-partite encoded TPS.
% The details are left to a future publication.

\textit{Example 6: tri-partite hybrid TPS}.--- Let us exhibit an
unusual example, of a TPS wherein each factor is of a different
physical nature. We consider $\mathcal{H}:=(\CC^{2})^{\otimes \,4}$
and $\mathcal{B}_{1}=\openone \otimes \mbox{End}(\CC^{2})^{\otimes
\,3}$ (full operator space over the last three qubits),
$\mathcal{B}_{2}=\openone \otimes \CC\mathcal{S}_{3}$ (permutations
exchanging the last three qubits). $\mathcal{B}_{1}$ is a factor and
one obtains the decomposition $ \mathcal{H}=\CC^{2}\otimes
\CC^{8}$. The three-qubit space splits wrt $\mathcal{S}_{3}$ as
$\CC^{4}\otimes \CC\oplus \CC ^{2}\otimes \CC^{2}$. It follows that
$(\CC^{2})^{\otimes \,4}\cong \CC ^{2}\otimes \CC^{4}\otimes \CC\oplus
\CC^{2}\otimes \CC^{2}\otimes \CC^{2}$.  The last term corresponds to
a \emph{tri-partite} system in which the first subsystem is a
``standard'' qubit, the second is acted upon by collective
interactions over the last three ``physical'' qubits, while the third
is acted upon by the algebra of permutations of $\mathcal{S}_{3}$.
Interestingly, this \emph{hybrid} tri-partite system has already been
realized experimentally in the context of noiseless-subsystems
\cite{vio-sci}.

%%% \textit{Example 7: An exception}.--- Not every subalgebra chain
%%% gives rise to a TPS. Consider the chain $\mathcal{S}_{N}\supset
%%% \mathcal{S}_{N-1}\supset ...\supset \mathcal{S}_{1}$ acting on
%%% $\mathcal{H}_{N}\cong $ $(\CC^{2})^{\otimes N}$, such that
%%% $\mathcal{S}_{N-i}$ has trivial action over the first $i$
%%% qubits. Using the notation of Ex.~5, we have
%%% $\mathcal{H}_{N-1}\cong \oplus _{J}H_{J}\otimes
%%% \CC^{\tilde{d}_{J}}$ and find ($J>0$) \cite{calc2}:
%%% $\CC^{\tilde{d}_{J}}\cong \CC^{\tilde{d}_{K-1/2}}\oplus
%%% \CC^{\tilde{d}_{K+1/2}}$. These two terms correspond to
%%% \emph{inequivalent} $\mathcal{S}_{N-1}$ irreps, hence there is no
%%% non-trivial TPS [the condition ``at least one $n_{J_{k}}>1$'',
%%% following Eq.~(\ref{chain-split}), is violated].

\textit{TPS morphing}.--- So far we have emphasized kinematics. Next we
show that an induced TPS can change dynamically, depending on the
algebras of available interactions. Let
$\{\mathcal{A}_{i}\}_{i=1}^{n}$ and
$\{\tilde{\mathcal{A}}_{i}\}_{i=1}^{\tilde{n}}$ define two TPSs over $
\mathcal{H}.$ Suppose one has the following Hamiltonian
\begin{equation}
 H(\lambda ,\mu )=\sum_{i=1}^{n}\sum_{\alpha }\lambda _{i}^{\alpha }H_{\alpha
 }^{i}+\sum_{i=1}^{\tilde{n}}\sum_{\beta }\mu _{i}^{\beta }{\tilde{H}}
 _{\beta }^{i}  \label{morph}
\end{equation}
where $H_{\alpha }^{i}\in \mathcal{A}_{i},\,{\tilde{H}}_{\beta
}^{i}\in \tilde{\mathcal{A}}_{i},\,(i=1,\ldots ,n)$, and all coupling
constants $\lambda _{i}$, $\mu _{i}$ are independently tunable.
%% and that, for any $i$, the set of Hamiltonians $\{H_{\alpha
%% }^{i}\}_{\alpha }$ ($ \{{\tilde{H}}_{\beta }^{i}\}_{\beta }$)
%% allows for universal control of the $i$-th factor of the TPS
%% induced by $\{\mathcal{A}_{i}\}_{i=1}^{n}$
%% ($\{\tilde{\mathcal{A}}_{i}\}_{i=1}^{\tilde{n}}$).
By setting all the $\mu _{i}$ ($\lambda _{i}$)
to zero the first (second) TPS is induced. Therefore, dynamical
control of the Hamiltonian allows to switch among different induced
multi-partitions, possibly with a different number of subsystems, in a
sort of continuous fashion. We call this ``TPS \emph{morphing}''. For
example, consider three qubits with controllable Hamiltonian given by
$H(\lambda (t),\mu (t))=\sum_{\imath ,j=1}^{3}\lambda
_{1}^{ij}\bm{\sigma }_{i}\cdot \bm{\sigma }_{j}+\sum_{\alpha
=x,y,z}\lambda _{2}^{\alpha }S^{\alpha }+\sum_{i=1}^{3}\sum_{\beta
=x,y,z}\mu _{i}^{\beta }\sigma _{i}^{\beta },$ where $S^{\alpha
}=\sum_{i=1}^{3}\sigma _{i}^{\alpha }\,(\alpha =x,y,z)$. The first two
terms induce the (encoded) \emph{bi-partite} TPS described in Ex.~2,
whereas the last term induces the standard \emph{tri-partite}
structure.

\textit{Stroboscopic entanglement}.---
%% Not only is a TPS a dynamical concept in the ``morphing'' sense,
A TPS can even be switched on and off under appropriate
circumstances. Suppose that the algebra of available interactions does
not induce a TPS [e.g., since it is $\cong
\mathrm{End}(\mathcal{H})$]. Now suppose that one can turn on an
additional interaction that allows one to refocus (see, e.g.,
\cite{LW}) some of these interactions, so that the remaining
interactions do induce a TPS. Then at the end of each refocusing
period a TPS will appear.  We call this ``\emph{stroboscopic
entanglement}''. For instance, and referring back to Ex.~1, suppose
that the controllable Hamiltonian is given by $H=\sum_{X\in
\mathcal{A}_{\chi },Y\in \mathcal{A}_{\lambda }}J_{X}X+J_{Y}Y$, where
the two-body terms are always on and the one-body terms are
controllable. This $H $ mixes the subalgebras $\mathcal{A}_{\chi }$
and $\mathcal{A}_{\lambda }$, so that there is no TPS as long as the
two-body terms are present. However, a series of $\pi $-pulses in
terms of $\sigma ^{x}\otimes \openone$ ($ \openone\otimes \sigma
^{z}$) will refocus, i.e., turn off, the two-body terms in
$\mathcal{A}_{\chi }$ ($\mathcal{A}_{\lambda }$) term, thus decoupling
the two subalgebras at the end of each refocusing period. In this
manner the $\chi $ and $\lambda $ factors can be separately
manipulated, i.e., the TPS has reappeared.

\textit{Conclusions}.--- We have shown that the TPS of quantum
mechanics acquires physical meaning only relative to the given set of
available interactions and measurements. These induce a TPS through their
algebraic structure. The induced TPS may contain factors (``qudits'')
of a different physical nature, and can be dynamical.

A few concluding comments are in order. First, note that while we
have given criteria for the appearance of an induced TPS and the
associated entanglement, we have deliberately {\em not} addressed the issue of {\em
  efficiency} in quantum information processing  (QIP) \cite{BD}, in particular in
relation to the question of {\em resource cost}. Indeed, it is simple
to construct a set of subalgebras satisying axioms i)-iii), thus
inducing a TPS for a ``structureless'' Hilbert space such
as energy levels of a Rydberg atom, while the associated cost of
performing a quantum computation scales exponentially in some resource
such as spectroscopic resolution \cite{Meyer}. Second, and again in the context
of QIP, in order
to exploit a given induced TPS for performing quantum computation one has to be able to
implement, along with the local operations $\mathcal{A}_{i}$, at least
one entangling transformation $\mathcal{E}$ in End$( \mathcal{C})\cong
\otimes_{i}\mathcal{A}_{i}$ \cite{uni}. The new set
$\{\{\mathcal{A}_{i}\},\mathcal{E}\}$, in the prototypical situation of
interest in QIP, will be (encoded) universal, i.e., will allow any
transformation in End$( \mathcal{C}$) to be {\em generated} by
composition of elementary operations involving
$\{\{\mathcal{A}_{i}\},\mathcal{E}\}$. This will allow access to other TPSs
than the original, induced one (e.g., in the case of Ex.~0 one could
argue that access to both the standard and the $\chi,\lambda$
bipartitions is available once all $SU(4)$ transformations can be
generated). The key point is that there is a {\em hierarchy} of TPSs:
the ``natural'' one is the one that is induced by the {\em directly}
accessible observables $\mathcal{A}_{i}$. The ``lower-level'' ones are those that are
visible only by composition of the elementary observables $\{\{\mathcal{A}_{i}\},\mathcal{E}\}$.
Third, it is important to emphasize that {\em both} interactions
and measurements are involved in inducing a TPS, and must be
{\em compatible}, i.e., induce the same TPS, for this TPS to be both
manipulable and observable.

%Finally, we note that exisiting measures of entanglement, as far as we
%know, make no reference to the notion of accessible
%interactions. Hence, an open question
%for future research is to devise a corresponding entanglement measure.

P.Z. gratefully acknowledges financial support by
Cambridge-MIT Institute Limited and by the European Union project  TOPQIP
(Contract IST-2001-39215). D.A.L. gratefully acknowledges the Alfred P. Sloan Foundation for a
Research Fellowship.


\begin{references}
\bibitem{proof}
Consider the subalgebras ${\cal A}_{1}^{(i)}:={\cal A} _{i}$ and
${\cal A}_{2}^{(i)}:=\vee _{j=i+1}^{n}{\cal A}_{j}\,(i=1,\ldots
,n-1)$. Let $i=1$. From iii) we have ${\cal A}_{1}^{(1)}\vee {\cal
A}_{2}^{(1)}\cong {\cal A}_{1}^{(1)}\otimes {\cal A}_{2}^{(1)}\cong
{\rm End}({\cal C})$. Consider the map $ \iota \colon {\cal
A}_{1}^{(1)}\otimes {\cal A}_{2}^{(1)}\rightarrow {\cal A}
_{1}^{(1)}\vee {\cal A}_{2}^{(1)}\colon x\otimes y\rightarrow
xy$. This is clearly a surjective (onto) algebra homomorphism. In
order to have ${\cal A} _{1}^{(1)}\vee {\cal A}_{2}^{(1)}\cong {\cal
A}_{1}^{(1)}\otimes {\cal A} _{2}^{(1)}$ it is necessary that
Ker$(\iota )=0$. In turns this implies: (A) ${\cal A}_{1}^{(1)}\cap
{\cal A}_{2}^{(1)}=\CC\openone$. Indeed if one had a non-trivial $x\in
{\cal A}_{1}^{(1)}\cap {\cal A}_{2}^{(1)}$ then the element $x\otimes
\openone-\openone\otimes x$ would be in Ker$(\iota)$.  Now we consider
the irrep decomposition of ${\cal A}_{1}^{(1)}$, i.e., ${\cal
A}_{1}^{(1)}\cong \oplus _{J}\openone_{n_{J}}\otimes M_{d_{J}}$; since
$ [{\cal A}_{2}^{(1)},{\cal A}_{1}^{(1)}]$, ${\cal A}_{2}^{(1)}\in
[{\cal A}_{1}^{(1)}]'\cong \oplus _{J}M_{n_{J}}\otimes
\openone_{d_{J}}$. Then ${\cal A}_{1}^{(1)}\cap {\cal
A}_{2}^{(1)}\subset \oplus _{J}\openone_{n_{J}}\otimes \openone
_{d_{J}}$. The only decomposition compatible with (A) is when just one
irrep appears, i.e., ${\cal A}_{1}^{(1)}\cong M_{d_{J}}\otimes
\openone_{n_{J}}$.  Since ${\cal A}_{1}^{(1)}\vee {\cal
A}_{1}^{(2)}\cong \mbox{End}({\cal H} )\cong M_{n_{J}d_{J}}$ one finds
that ${\cal A}_{1}^{(1)}$ must be isomorphic to $M_{n_{J}}$. In
summary: $\exists {\cal H}_{1}^{(i)}\cong \CC^{d_{J}},\,{\cal
H}_{2}^{(i)}\cong \CC^{n_{J}}$ such that ${\cal C}\cong \otimes
_{l=1}^{2}{\cal H}_{l}^{(1)}$ and ${\cal A}_{l}^{(1)}\cong \mbox{End}
({\cal H}_{l}^{(1)}),\,(l=1,2)$. This argument can be iterated over
$i$ for $ {\cal A}_{1}^{(i)}$ and ${\cal A}_{2}^{(i)}$. The role of
${\cal C}$ is played by the second state-factor obtained for
$i-1$. Eventually one finds $ {\cal C}\cong \otimes _{i=1}^{n}{\cal
H}_{1}^{(i)},$ with ${\cal A}_{i}\cong \mbox{End}({\cal
H}_{1}^{(i)})$. $\hfill\Box$
\bibitem{proof2}
Upon restriction to a particular code subspace we have $
\mathcal{B}_{i-1}\cong \otimes _{k=1}^{i-1}\openone_{n_{J_{k}}}\otimes
M_{d_{_{J_{i-1}}}}$ and $\mathcal{B}_{i}^{\prime }\cong \otimes
_{k=1}^{i}M_{n_{J_{k}}}\otimes \openone_{d_{J_{i}}}$. Then for $i<n$:
$\mathcal{B} _{i}^{\prime }\cap \mathcal{B}_{i-1}=\otimes
_{k=1}^{i-1}\openone _{n_{J_{k}}}\otimes M_{n_{J_{i}}}\otimes
_{k=i+1}^{n}\openone _{n_{J_{k}}}\otimes \openone_{J_{d_{n}}}$, i.e.,
has non-trivial action only on the $i$th subsystem. For $i=n$ we have
$\mathcal{B}_{n}^{\prime }\cap \mathcal{B}_{n-1}=\otimes
_{k=1}^{n}\openone_{n_{J_{k}}}\otimes M_{d_{J_{n}}}$. For the converse
direction, the formula $\mathcal{B} _{i}:=\vee
_{k=i+1}^{n}\mathcal{A}_{k}$ is immediate. $\hfill\Box$
\bibitem{calc}
We have to decompose the $\CC^{\tilde{d}_{J}}$ factors wrt
$\mathcal{S}_{3}\times \mathcal{S}_{3}$. Using the $ \mathcal{S}_{3}$
decomposition $\mathcal{H}_{3}\cong H_{1/2}\otimes \CC ^{2}\oplus
H_{3/2}\otimes \CC$, we have $\mathcal{H}_{6}\cong (\mathcal{H}
_{3})^{\otimes 2}\cong (H_{1/2})^{\otimes 2}\otimes
\CC^{4}\oplus...$. Using the su$(2)$ result $H_{J_{1}}\otimes
H_{J_{2}}\cong \oplus _{|J_{1}-J_{2}|\leq J\leq J_{1}+J_{2}}H_{J}$ we
can collect together the terms in the expansion of
$(\mathcal{H}_{3})^{\otimes 2}$ to match those in $
\mathcal{H}_{6}$. Focusing on the $\CC^{\tilde{n}_{J_{1}}}\otimes
\CC^{ \tilde{d}_{J_{1}}}=H_{1}\otimes \CC^{9}$ term in the
decomposition of ${\cal H}_6$, we find that it splits as $H_{1}\otimes
(\CC\oplus \CC^{2}\oplus \CC^{2}\oplus \CC^{4})$. The middle two terms
represent a two-dimensional $\mathcal{S} _{3}\times
\mathcal{S}_{3}$-irrep appearing with multiplicity two, and can be
then written as $\CC^{\tilde{n}_{J_{2}}}\otimes
\CC^{\tilde{d}_{J_{2}}}= \CC^{2}\otimes \CC^{2}$. The claimed result
then follows.
\end{references}
\end{document}